\documentclass[
reprint,
 amsmath,amssymb,
 aps,
]{revtex4-1}

\usepackage{graphicx}
\usepackage{dcolumn}
\usepackage{bm}

\begin{document}

\preprint{APS/123-QED}

\title{Local Spin Susceptibility of the $S=1/2$ Kagome Lattice in ZnCu$_{3}$(OD)$_{6}$Cl$_{2}$}

\author{T. Imai,$^{1,2}$ M. Fu,$^{1}$ T. H. Han,$^{3}$ and Y. S. Lee$^{3}$}
\affiliation{%
$^{1}$Department of Physics and Astronomy, McMaster University, Hamilton, Ontario L8S 4M1, Canada
}%
\affiliation{%
$^{2}$Canadian Institute for Advanced Research, Toronto M5G 1Z8, Canada
}%
\affiliation{%
$^{3}$Department of Physics, M.I.T., Cambridge, Massachusetts 02139, USA
}%

\date{\today}

\begin{abstract}
We report single-crystal $^{2}$D NMR investigation of the nearly ideal spin $S=1/2$ kagome lattice ZnCu$_{3}$(OD)$_{6}$Cl$_{2}$.  We successfully identify $^{2}$D NMR signals originating from the nearest-neighbors of Cu$^{2+}$ defects occupying Zn sites.  From the $^{2}$D Knight shift measurements, we demonstrate that weakly interacting Cu$^{2+}$ spins at these defects cause the large Curie-Weiss enhancement toward $T=0$ commonly observed in the bulk susceptibility data.  We estimate the intrinsic spin susceptibility of the kagome planes by subtracting defect contributions, and explore several scenarios.
\end{abstract}

\pacs{75.10.Kt, 76. 60-k}
\keywords{kagome, NMR, ZnCu$_{3}$(OH)$_{6}$Cl$_{2}$}
\maketitle

Persistent search for frustrated spin-liquid systems without a magnetically ordered ground state \cite{PhysicsToday, Lee} has recently resulted in the discovery of several model materials with spin $S=1/2$, including the kagome \cite{Shores,Helton,Hiroi}, triangular \cite{Kanoda}, and FCC \cite{Aharen, deVries1} lattice systems.  Intense experimental and theoretical efforts are underway to understand their ground state properties, and to identify potential signatures of exotic spin-liquid states.  Among the most promising candidates of spin-liquids is herbertsmithite ZnCu$_{3}$(OH)$_{6}$Cl$_{2}$, which contains planes of a kagome lattice formed by Cu$^{2+}$ ions with $S=1/2$ (see Fig.\ 1 for the crystal structure) \cite{Shores, Helton, Mendels, SHLee, Mendels, Imai, Olariu, deVries2, Han}.     The antiferromagnetic Cu-Cu exchange interaction $J$ is estimated to be as large as $J/k_{B} \sim 190$~K \cite{Helton},  yet there is a firm consensus that no hint of magnetic long range order exists at least down to 50~mK \cite{Helton,Mendels}. 

Nonetheless, the mechanism behind the realization of such a non-magnetic ground state in ZnCu$_{3}$(OH)$_{6}$Cl$_{2}$ has been controversial (see \cite{MendelsReview} for a recent review).  The root cause of the controversy stems from the difficulty in pinpointing the location of defects, and understanding their potential influence on various physical properties.   For example, magnetic Cu$^{2+}$ defects occupying the non-magnetic Zn$^{2+}$ sites outside the kagome plane might induce a large Curie-Weiss contribution $\chi_{CW}$ to the bulk-averaged spin susceptibility $\chi_{bulk}$, and mask its intrinsic behavior.  In fact, $\chi_{bulk}$ always exhibits a large enhancement below about 50~K, as shown in Fig.\ 2a.  The observed behavior is different from the predicted behavior of $\chi_{kagome}$ for ideal kagome antiferromagnets with vanishing spin susceptibility at $T=0$ \cite{Elstner,Mila,Rigol,Misguich,Ran,Hermele,Yan}.  Furthermore, if non-magnetic Zn$^{2+}$ defects occupy the magnetic Cu$^{2+}$ sites within the kagome plane, spin singlets may be induced in their vicinity toward $T=0$ \cite{Olariu}, and help realize the non-magnetic ground state.   These uncertainties over the nature of defects prevented extraction of the intrinsic $\chi_{kagome}$ from $\chi_{bulk}$.  It is therefore necessary to untangle the effects of defects based on real space probes. 

\begin{figure}[b]
\includegraphics[width=2.7in]{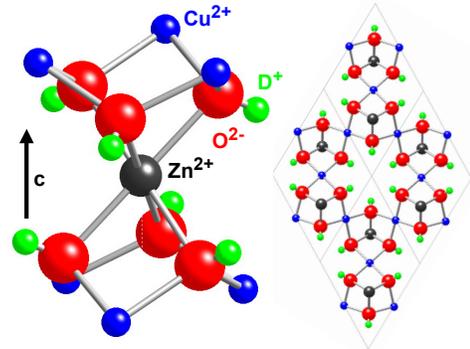}\\
\caption{\label{Fig.1} (Color Online) Left: local configuration of ions surrounding the Zn$^{2+}$ site in ZnCu$_{3}$(OD)$_{6}$Cl$_{2}$.  Right: the c-axis view of the kagome plane.
}
\end{figure}
\begin{figure}[b]
\includegraphics[width=3.2in]{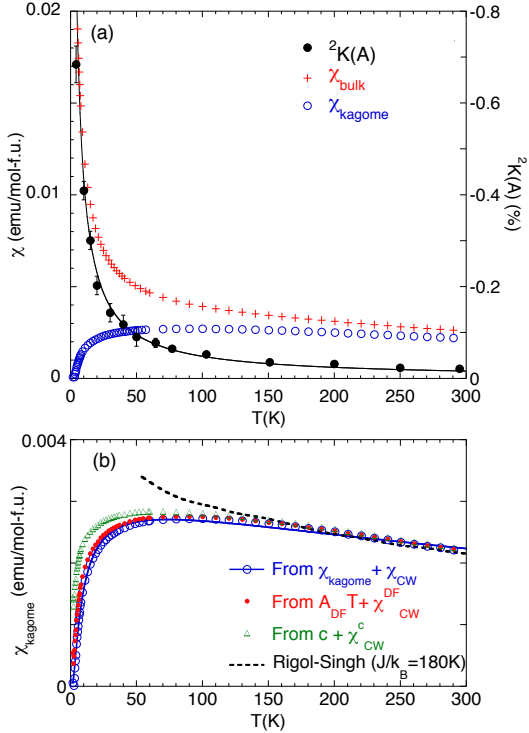}\\
\caption{\label{Fig.2} (Color Online) (a) ($+$): The bulk averaged magnetic susceptibility $\chi_{bulk}$ measured by SQUID in 0.2~Tesla for a collection of  small single crystals with random orientations.  The solid curve represents extrapolation of the Curie-Weiss contribution $\chi_{CW}$ with $\theta_{bulk}=-1.2$~K estimated by fitting $\chi_{bulk}$ below 3~K.  ($\bullet$): $^{2}$D NMR Knight shift $^{2}K(A)$ measured at the A site, which exhibits nearly identical temperature dependence as $\chi_{CW}$.  ($\circ$): The intrinsic spin susceptibility within the kagome plane, $\chi_{kagome}=\chi_{bulk}-\chi_{CW}$.  (b) ($\circ$): the same $\chi_{kagome}$ plotted in (a).  The solid curve through the data points represents the best fit with a small gap (see main text).  The dotted curve shows the theoretical prediction \cite{Rigol} with $J/k_{B}=180$~K.  ($\bullet$): $\chi_{kagome}^{DF}$ estimated from the best fit of $\chi_{bulk}$ below 8~K to a summation of the $T$-linear term from Dirac Fermions and a Curie-Weiss term.  ($\triangle$): an alternative scenario of $\chi_{kagome}^{c}$ with a finite gapless excitations $c$ at $T=0$.  
}
\end{figure}

In this Rapid Communication, we take advantage of recent successful growth of high quality single crystals of deuterized ZnCu$_{3}$(OD)$_{6}$Cl$_{2}$ \cite{Han}, and report a single-crystal NMR investigation.  Thanks to very sharp NMR lines in a 45~mg single crystal, we identified $^{2}$D NMR signals originating from the 6 nearest-neighbor (n.n.) and 12 next-nearest-neighbor (n.n.n.) $^{2}$D sites of Cu$^{2+}$ defects occupying the Zn sites.  From their relative intensities, we show that magnetic Cu$^{2+}$ defects replace non-magnetic Zn$^{2+}$ with a probability of $14\pm2$\%.   A Curie-Weiss behavior of the NMR Knight shift $^{2}K$ observed at the n.n. sites indicates that the large Curie-Weiss enhancement commonly observed for $\chi_{bulk}$ below $\sim 50$~K arises from these defects. By subtracting the contribution of the Curie-Weiss contribution $\chi_{CW}$  from $\chi_{bulk}$, we explore the intrinsic behavior of spin susceptibility of the kagome planes, $\chi_{kagome}$. 

In Fig.\ 3, we present representative $^{2}$D NMR lineshapes observed in an external magnetic field of $B_{ext}=8.4$~Tesla applied along the c-axis, {\it i.e.} normal to the kagome plane.  The $^{2}$D nucleus has a nuclear spin $I=1$, and hence we would expect to observe a pair of NMR transitions between the nuclear spin state $I_{z}=\pm1$ and $I_{z}=0$, {\it if there was no defect}.  In general, one can express the resonant frequency of each $^{2}$D site as $f_{\pm} = \gamma_{n} B_{total} \pm \nu_{Q}^{c}$, where the nuclear gyromagnetic ratio is $\gamma_{n}/2\pi = 6.53566$~MHz/Tesla, and $B_{total}$ represents the total magnetic field seen by each nuclear spin.  The second term, $\pm \nu_{Q}^{c}$, represents contributions of the c-axis component of the nuclear quadrupole interaction for the aforementioned $I_{z}=\pm1$ to $I_{z}=0$ transitions.  The $\nu_{Q}^{c}$ is proportional to the second derivative of the total Coulomb potential at the $^{2}$D sites, and small differences in the local structure lead to distinct values of the NMR line splitting $2\nu_{Q}^{c}$.  

In reality, we observe three pairs of $^{2}$D NMR signals, A$_{\pm}$, B$_{\pm}$, and C$_{\pm}$.  This means that there are three distinct $^{2}$D sites in ZnCu$_{3}$(OD)$_{6}$Cl$_{2}$  with locally different environment.  Based on the relative intensity of well-resolved peaks A$_{+}$, B$_{+}$, and C$_{+}$ at 295~K, we estimate the population of A, B, and C sites as $14\pm2$\%, $28\pm4$\%, and $58\pm4$\%, respectively.   From the line splitting, we obtain $\nu_{Q}^{c}=(f_{+}-f_{-})/2 = 48$, 62, and 45~kHz for  A$_{\pm}$, B$_{\pm}$, and C$_{\pm}$ peaks, respectively.  Within experimental uncertainties, we didn't observe any noticeable temperature dependence in $\nu_{Q}^{c}$.  Thus the temperature dependence of $f_{\pm}$ arises from magnetic effects through $\gamma_{n} B_{total}$.
\begin{figure}[b]
\includegraphics[width=3in]{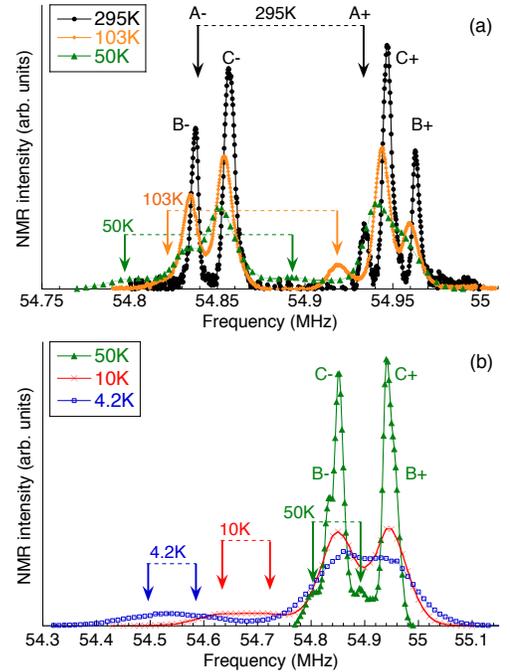}\\
\caption{\label{Fig.2} (Color Online) Representative $^{2}$D NMR lineshapes observed in $B_{ext}=8.4$~Tesla applied along the c-axis.  We observe three pairs of quandrupole-split NMR lines, A$_{\pm}$, B$_{\pm}$, and C$_{\pm}$.  Downward arrows mark the resonant frequencies $f_{\pm}$ for A$_{\pm}$.    Notice that the horizontal scale is changed for (b).  
}
\end{figure}

The $B_{total}$ at each $^{2}$D site is a summation of the external field $B_{ext}=8.4$~Tesla and the hyperfine magnetic field $B_{hf}$ from nearby electron spins polarized by $B_{ext}$, {\it i.e.} $B_{total}=B_{ext}+B_{hf}$.  Since  $B_{hf}$ is proportional to {\it local} spin susceptibility $\chi_{spin}$ in the vicinity of the observed $^{2}$D site, one can gain insight into the position dependent $\chi_{spin}$ within the kagome plane by keeping track with the temperature dependence of $f_{\pm}$ at different sites.  A striking feature of Fig.\ 3 is that A$_{\pm}$ peaks progressively shift to lower frequencies, and split off below 10~K.  As shown below, this is because A sites are under the direct influence of the Cu$^{2+}$ defect spins at n.n. Zn sites.  

In Fig.\ 4, we summarize the temperature dependence of the $^{2}$D NMR Knight shift $^{2}K = B_{hf}/B_{ext}=(B_{total}-B_{ext})/B_{ext}$, which is related to local spin susceptibility as $^{2}K = \Sigma_{i} A_{hf}^{(i)}\chi_{spin}^{(i)} + K_{chem}$.  $A_{hf}^{(i)}$ is the hyperfine coupling constant between the observed $^{2}$D nuclear spin and the electron spin at the $i$th Cu site in their vicinity, and $\chi_{spin}^{(i)}$ is the local spin susceptibility at the $i$th Cu site.  $A_{hf}^{(i)}$ is negative in the present case, hence we reverse the vertical axis of Fig.\ 4.  $K_{chem}$ is a small temperature independent chemical shift originating from the orbital motion of electrons.   The results in Fig.\ 4 indicate that the Knight shift $^{2}K(A)$ at the A sites is anomalously large compared with $^{2}K(B)$ and $^{2}K(C)$ at B and C sites.  Furthermore, the temperature dependence of $^{2}K(A)$ obeys a single Curie-Weiss law in the entire temperature range with a Weiss temperature, $\theta_{A}=-3.5$~K.  

In Fig.\ 2, we compare the temperature dependences of $^{2}K(A)$ and $\chi_{bulk}$.  It is interesting to note that $^{2}K(A)$ asymptotes to $\chi_{bulk}$ at the low temperature limit.  In fact, the low temperature behavior of $\chi_{bulk}$ can be approximated by a Curie-Weiss law $\chi_{CW} \sim 0.12/(T+\theta_{bulk})$~emu/mol-f.u. with a vanishingly small Weiss temperature, $\theta_{bulk}=-1.2$~K; the extrapolation of $\chi_{CW}$ to higher temperature reproduces the temperature dependence of $^{2}K(A)$, as shown by a solid curve.  The slight overestimation of $\theta_{A}=-3.5$~K in $B_{ext}=8.4$~Tesla by NMR over $\theta_{bulk}=-1.2$~K should be attributed to a saturation of defect spin polarization in high fields below $T\sim \mu_{B}B_{ext}/k_{B} \sim 5.6$~K.  We confirmed that $^{2}K(A)$ at 4.2~K measured in $B_{ext}=5.3$ and 3.2~Tesla is indeed somewhat larger, and the field dependence is consistent with that of bulk magnetization.  

\begin{figure}[b]
\includegraphics[width=3.2in]{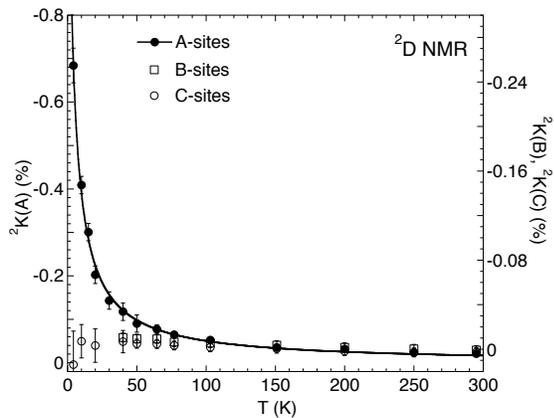}\\
\caption{\label{Fig.4} NMR Knight shift $^{2}K$ at three distinct $^{2}$D sites.  Solid curve represents the best Curie-Weiss fit.
}
\end{figure}

Thus we have identified the source of the large enhancement of $\chi_{bulk}$ observed below 50 K as arising from defect moments neighboring $^{2}$D A sites whose behavior can be modeled with a Curie-Weiss law.  The vanishingly small Weiss temperature implies that interactions among these defect moments are very weak.  This weak coupling may explain the effectiveness of a simple Curie-Weiss law, even though a distribution of coupling strengths would be expected for a random arrangement of defects.  The next question to address then, is: where are these defects located?  We recall that A sites account for $14\pm2$\% of the overall $^{2}$D NMR signals, and the intensity of B sites is precisely twice greater, $28\pm4$\%.  In view of the fact that each Cu$^{2+}$ defect occupying the Zn sites has 6 n.n. and 12 n.n.n. $^{2}$D sites (see Fig.\ 1), we assign A and B sites as the n.n. and n.n.n. sites of the Cu$^{2+}$ defects, respectively.  These site assignments also imply that $14\pm2$\% of the Zn sites in our single crystal are occupied by weakly interacting Cu$^{2+}$ spins, and there is no $^{2}$D NMR signature of Zn$^{2+}$ anti-site defects occupying the Cu$^{2+}$ kagome sites.  Our findings are consistent with the recent site-specific x-ray anomalous dispersion measurements combined with Rietvelt refinement, from which the actual composition of our sample was estimated as (Zn$_{0.85}$Cu$_{0.15}$)Cu$_{3}$(OD)$_{6}$Cl$_{2}$ \cite{Freedman}.    

We gain additional insight into the nature of defects from the sign of the hyperfine interaction, $A_{hf}$.  Our observation, $^{2}K(A) < 0$, implies that the hyperfine field $B_{hf}$ at $^{2}$D A sites, induced by the defect moments polarized along the c-axis in $B_{ext}$, is {\it negative}.  From the comparison of the temperature dependence of $^{2}K(A)$ and $\chi_{CW}$, we estimate $A_{hf} \propto d(^{2}K)/d(\chi_{CW}) = - 24$~kOe/$\mu_{B}$.  On the other hand, in the present geometry employed for our NMR measurements, it is easy to show that the dipole hyperfine field from Zn sites, $A_{hf}^{dipole} = + 0.85$~kOe/$\mu_{B}$, has a {\it positive} component along the c-axis.  Fermi's contact hyperfine interaction induced by direct overlap of the $^{2}$D 1s orbital and Cu 3d$^{9}$ orbitals at Zn sites is also {\it positive}.  These considerations suggest that a sizable spin polarization obeying the same Curie-Weiss behavior also exists in the Cu$_{3}$(OD)$_{3}$ triangles within the kagome planes above and below the Zn defect sites; the negative hyperfine field at $^{2}$D A sites originates from these spin polarizations.  This conclusion also explains the fact that $14\pm2$\% of isolated Cu$^{2+}$ defects alone would underestimate the magnitude of $\chi_{CW}$ by a factor of $\sim2$.

In contrast with the Curie-Weiss behavior of A sites, the Knight shift $^{2}K(C)$ at C sites is very small, and doesn't show a large enhancement toward $T=0$, as shown in Fig\ 4.  In other words, at least 58\% of Cu sites within the kagome planes  show no hint of enhancement of local spin susceptibility $\chi_{spin}$ toward $T=0$.  This means that the defect moments do not significantly couple to the Cu ions in the kagome plane beyond the n.n.n. Cu sites, with which the 2D C sites are bonded.  Since we can't resolve the $^{2}$D B sites from C sites below$\sim 50$~K, 28\% of Cu sites adjacent to B sites don't have a large enhancement of $\chi_{spin}$, either, although we can't entirely rule out a mild enhancement.   

We have established that the Curie-Weiss contribution $\chi_{CW}$ to $\chi_{bulk}$ originates from the immediate vicinity of the Cu$^{2+}$ defect moments occupying the Zn sites.  This finding paves a new avenue to analyze the intrinsic spin susceptibility within the kagome plane, $\chi_{kagome}$, by subtracting $\chi_{CW}$ from $\chi_{bulk}$, {\it i.e.}  $\chi_{kagome} = \chi_{bulk}-\chi_{CW}$.  In earlier studies of various quantum antiferromagnets, $\chi_{CW}$ was generally subtracted from $\chi_{bulk}$ to reveal intrinsic behavior, {\it e.g.} \cite{Azuma}.  Because of the uncertainties in the origin of defects, such analysis of  $\chi_{kagome}$ has not been fully explored in the present case.  In Fig.\ 2(a), we present $\chi_{kagome} = \chi_{bulk}-\chi_{CW}$, together with $\chi_{bulk}$ and $\chi_{CW}$.  We also present the same $\chi_{kagome}$ in Fig.\ 2(b) in an expanded scale.

Once we subtract $\chi_{CW}$, a qualitatively different behavior of $\chi_{kagome}$ emerges; $\chi_{kagome}$ saturates below $\sim 100$~K, then decreases with temperature toward $T=0$.  The temperature dependence of $\chi_{kagome}$ is consistent with the theoretical prediction based on a high temperature series expansion with $J/k_{B}\sim180$~K as shown by a dotted curve. Our finding near $T=0$ is also consistent with earlier reports based on  $^{35}$Cl and $^{17}$O powder NMR that $\chi_{spin}$ at some fraction of Cu sites decreases toward $T=0$ \cite{Imai, Olariu}.  However, dramatic NMR line broadening hampered the efforts to probe the bulk-averaged behavior below 50~K in these earlier powder NMR studies.

The solid curve in Fig.\ 2b represents the best fit to an empirical formula $\chi_{kagome}= A \cdot exp(- \Delta/k_{B}T)/(T+\theta_{kagome})$ with a small gap $\Delta/k_{B}\sim 7$~K and $\theta_{kagome} \sim 690$~K.  The fit is reasonably good, but we should be cautious in interpreting its implications.  Since we estimated $\chi_{CW}$ from $\chi_{bulk}$ below 3~K, and subtracted the former from the latter, in essence, we are forcing  $\chi_{kagome}$ to reach zero at 3~K.  In other words, the estimated values of $\chi_{kagome}$ set the lower bound of the intrinsic spin contributions at kagome sites. Thus we can't rule out the possibility that this small gap is an artifact of our analysis.  In this sense, we should consider $\Delta\sim 0.04J$ as the {\it upper bound} of any possible spin gap.  

In fact, we noticed that the temperature dependence of  $\chi_{kagome}$ near $T=0$ is also consistent with a $T$-linear behavior predicted for gapless excitations of the Dirac Fermion in algebraic spin liquids, $\chi_{kagome}= A_{DF}T$, where $A_{DF}=9.6 N_{A} \mu_{B}^{2} k_{B}/J^{2}$~emu/Kmol-f.u.\cite{Ran,Hermele}.  It is worth noting that the same Dirac Fermion model successfully accounts for the $C_{spin} \sim T^{2}$ behavior of the specific heat data below 30~K with $J/k_{B}\sim 190$~K \cite{Ran, Hermele}.  Inspired by these promising signatures, we also fitted $\chi_{bulk}$ as a summation of the Dirac Fermion contributions $\chi_{kagome}^{DF}= A_{DF}T$ and defect contributions $\chi_{CW}^{DF}$ below 8~K, and estimate $\chi_{kagome}^{DF} = \chi_{bulk}-\chi_{CW}^{DF}$.  It turned out that $\chi_{CW}^{DF}$ is nearly identical with the previous estimate, hence $\chi_{kagome}^{DF}$ shows very similar behavior, as shown in Fig.2b.  The best fitted value, $A_{DF} = 2.0\times 10^{-4}$~emu/K mol-f.u., implies $J/k_{B}\sim 220$~K.  This is in reasonable agreement with $J/k_{B}=180\sim 190$~K mentioned above.

While the predictions based on Dirac Fermions with zero density of states at the Fermi energy account for both $\chi_{kagome}$ and $C_{spin}$ fairly well, we caution that the nature of low lying states of spin liquids is still controversial.  For example, spin liquids with a spinon Fermi surface on triangular lattices are known to exhibit a finite density of states \cite{Kanoda, SSLee}.  An analogous scenario, however, seems highly unlikely in the present case, because $C_{spin}$ exhibits positive curvatures in a broad temperature range below 30~K \cite{Helton}, instead of a $T$-linear behavior expected for a finite density of states.  We estimate the upper bound of the $T$-linear term as $C_{spin}\sim0.1T$ J/mol-f.u. K$^{2}$ below 1~K in 14~Tesla \cite{Helton}.  This translates to the upper bound of the constant term in the $T=0$ limit of $\chi_{kagome}$ as $c=1.4\times 10^{-3}$ emu/mol-f.u.  We fitted the SQUID data to $\chi_{bulk} = c + \chi_{CW}^{c}$ below 3~K, and subtracted $\chi_{CW}^{c}$ from $\chi_{bulk}$  to estimate $\chi_{kagome}^{c}$ in the entire temperature range.  The obtained results in Fig.2b are similar to the previous two cases, except for the finite intercept at $T=0$.

To summarize, we identified the source of a large enhancement of spin susceptibility in ZnCu$_{3}$(OH)$_{6}$Cl$_{2}$ as weakly interacting defect moments induced by Cu$^{2+}$ in Zn sites and their immediate neighbors, and paved an avenue to analyze the bulk spin susceptibility data.  By subtracting the defect contributions, we estimated an upper and lower bound of the intrinsic spin susceptibility within the kagome planes, $\chi_{kagome}^{c}$ (with a finite intercept) and $\chi_{kagome}$ (with a small gap), respectively.  The reality must be somewhere in between.  

T.I. thanks E. Sorensen, S.-S. Lee, P.A. Lee, Y. Ran, D. Freedman, and H. Eisaki for helpful communications, and acknowledges the financial support by NSERC and CIFAR.  The work at MIT was supported by the US Department of Energy under Grant No. DE-FG02-07ER46134.


\begin{thebibliography}{25}%
\makeatletter
\providecommand \@ifxundefined [1]{%
 \@ifx{#1\undefined}
}%
\providecommand \@ifnum [1]{%
 \ifnum #1\expandafter \@firstoftwo
 \else \expandafter \@secondoftwo
 \fi
}%
\providecommand \@ifx [1]{%
 \ifx #1\expandafter \@firstoftwo
 \else \expandafter \@secondoftwo
 \fi
}%
\providecommand \natexlab [1]{#1}%
\providecommand \enquote  [1]{``#1''}%
\providecommand \bibnamefont  [1]{#1}%
\providecommand \bibfnamefont [1]{#1}%
\providecommand \citenamefont [1]{#1}%
\providecommand \href@noop [0]{\@secondoftwo}%
\providecommand \href [0]{\begingroup \@sanitize@url \@href}%
\providecommand \@href[1]{\@@startlink{#1}\@@href}%
\providecommand \@@href[1]{\endgroup#1\@@endlink}%
\providecommand \@sanitize@url [0]{\catcode `\\12\catcode `\$12\catcode
  `\&12\catcode `\#12\catcode `\^12\catcode `\_12\catcode `\%12\relax}%
\providecommand \@@startlink[1]{}%
\providecommand \@@endlink[0]{}%
\providecommand \url  [0]{\begingroup\@sanitize@url \@url }%
\providecommand \@url [1]{\endgroup\@href {#1}{\urlprefix }}%
\providecommand \urlprefix  [0]{URL }%
\providecommand \Eprint [0]{\href }%
\@ifxundefined \urlstyle {%
  \providecommand \doi  [0]{\begingroup \@sanitize@url \@doi}%
  \providecommand \@doi [1]{\endgroup \@@startlink {\doibase
  #1}doi:\discretionary {}{}{}#1\@@endlink }%
}{%
  \providecommand \doi  [0]{doi:\discretionary{}{}{}\begingroup
  \urlstyle{rm}\Url }%
}%
\providecommand \doibase [0]{http://dx.doi.org/}%
\providecommand \Doi [0]{\begingroup \@sanitize@url \@Doi }%
\providecommand \@Doi  [1]{\endgroup\@@startlink{\doibase#1}\@@Doi}%
\providecommand \@@Doi [1]{#1\@@endlink}%
\providecommand \selectlanguage [0]{\@gobble}%
\providecommand \bibinfo  [0]{\@secondoftwo}%
\providecommand \bibfield  [0]{\@secondoftwo}%
\providecommand \translation [1]{[#1]}%
\providecommand \BibitemOpen [0]{}%
\providecommand \bibitemStop [0]{}%
\providecommand \bibitemNoStop [0]{.\EOS\space}%
\providecommand \EOS [0]{\spacefactor3000\relax}%
\providecommand \BibitemShut  [1]{\csname bibitem#1\endcsname}%
\bibitem [{Phy(2007)}]{PhysicsToday}%
  \BibitemOpen
  \href@noop {} {\bibfield  {journal} {\bibinfo  {journal} {Physics Today},\
  }\textbf {\bibinfo {volume} {60}},\ \bibinfo {pages} {(2), 16} (\bibinfo
  {year} {2007})}\BibitemShut {NoStop}%
\bibitem [{\citenamefont {Lee}(2008)}]{Lee}%
  \BibitemOpen
  \bibfield  {author} {\bibinfo {author} {\bibfnamefont {P.~A.}\ \bibnamefont
  {Lee}},\ }\href@noop {} {\bibfield  {journal} {\bibinfo  {journal}
  {Science},\ }\textbf {\bibinfo {volume} {321}},\ \bibinfo {pages} {1306}
  (\bibinfo {year} {2008})}\BibitemShut {NoStop}%
\bibitem [{\citenamefont {Shores}\ \emph {et~al.}(2005)\citenamefont {Shores},
  \citenamefont {Nytko}, \citenamefont {Bartlett},\ and\ \citenamefont
  {Nocera}}]{Shores}%
  \BibitemOpen
  \bibfield  {author} {\bibinfo {author} {\bibfnamefont {M.~P.}\ \bibnamefont
  {Shores}}, \bibinfo {author} {\bibfnamefont {E.~A.}\ \bibnamefont {Nytko}},
  \bibinfo {author} {\bibfnamefont {B.~M.}\ \bibnamefont {Bartlett}}, \ and\
  \bibinfo {author} {\bibfnamefont {D.~G.}\ \bibnamefont {Nocera}},\
  }\href@noop {} {\bibfield  {journal} {\bibinfo  {journal} {J.\ Am.\ Chem.\
  Soc.},\ }\textbf {\bibinfo {volume} {127}},\ \bibinfo {pages} {13462}
  (\bibinfo {year} {2005})}\BibitemShut {NoStop}%
\bibitem [{\citenamefont {Helton}\ \emph {et~al.}(2007)\citenamefont {Helton},
  \citenamefont {Matan}, \citenamefont {Shores}, \citenamefont {Nytko},
  \citenamefont {Bartlett}, \citenamefont {Yoshida}, \citenamefont {Takano},
  \citenamefont {Suslov}, \citenamefont {Qiu}, \citenamefont {Chung},
  \citenamefont {Nocera},\ and\ \citenamefont {Lee}}]{Helton}%
  \BibitemOpen
  \bibfield  {author} {\bibinfo {author} {\bibfnamefont {J.~S.}\ \bibnamefont
  {Helton}}, \bibinfo {author} {\bibfnamefont {K.}~\bibnamefont {Matan}},
  \bibinfo {author} {\bibfnamefont {M.~P.}\ \bibnamefont {Shores}}, \bibinfo
  {author} {\bibfnamefont {E.~A.}\ \bibnamefont {Nytko}}, \bibinfo {author}
  {\bibfnamefont {B.~M.}\ \bibnamefont {Bartlett}}, \bibinfo {author}
  {\bibfnamefont {Y.}~\bibnamefont {Yoshida}}, \bibinfo {author} {\bibfnamefont
  {Y.}~\bibnamefont {Takano}}, \bibinfo {author} {\bibfnamefont
  {A.}~\bibnamefont {Suslov}}, \bibinfo {author} {\bibfnamefont
  {Y.}~\bibnamefont {Qiu}}, \bibinfo {author} {\bibfnamefont {J.~H.}\
  \bibnamefont {Chung}}, \bibinfo {author} {\bibfnamefont {D.~G.}\ \bibnamefont
  {Nocera}}, \ and\ \bibinfo {author} {\bibfnamefont {Y.~S.}\ \bibnamefont
  {Lee}},\ }\href@noop {} {\bibfield  {journal} {\bibinfo  {journal} {Phys.\
  Rev.\ Lett.},\ }\textbf {\bibinfo {volume} {98}},\ \bibinfo {pages} {107204}
  (\bibinfo {year} {2007})}\BibitemShut {NoStop}%
\bibitem [{\citenamefont {Hiroi}\ \emph {et~al.}(2001)\citenamefont {Hiroi},
  \citenamefont {Hanawa}, \citenamefont {Kobayashi}, \citenamefont {Nohara},
  \citenamefont {Takagi}, \citenamefont {Kato},\ and\ \citenamefont
  {Takigawa}}]{Hiroi}%
  \BibitemOpen
  \bibfield  {author} {\bibinfo {author} {\bibfnamefont {Z.}~\bibnamefont
  {Hiroi}}, \bibinfo {author} {\bibfnamefont {M.}~\bibnamefont {Hanawa}},
  \bibinfo {author} {\bibfnamefont {N.}~\bibnamefont {Kobayashi}}, \bibinfo
  {author} {\bibfnamefont {M.}~\bibnamefont {Nohara}}, \bibinfo {author}
  {\bibfnamefont {H.}~\bibnamefont {Takagi}}, \bibinfo {author} {\bibfnamefont
  {Y.}~\bibnamefont {Kato}}, \ and\ \bibinfo {author} {\bibfnamefont
  {M.}~\bibnamefont {Takigawa}},\ }\href@noop {} {\bibfield  {journal}
  {\bibinfo  {journal} {J. Phys. Soc. Jpn.},\ }\textbf {\bibinfo {volume}
  {70}},\ \bibinfo {pages} {3377} (\bibinfo {year} {2001})}\BibitemShut
  {NoStop}%
\bibitem [{\citenamefont {Shimizu}\ \emph {et~al.}(2003)\citenamefont
  {Shimizu}, \citenamefont {Miyagawa}, \citenamefont {Kanoda}, \citenamefont
  {Maesato},\ and\ \citenamefont {Saito}}]{Kanoda}%
  \BibitemOpen
  \bibfield  {author} {\bibinfo {author} {\bibfnamefont {Y.}~\bibnamefont
  {Shimizu}}, \bibinfo {author} {\bibfnamefont {K.}~\bibnamefont {Miyagawa}},
  \bibinfo {author} {\bibfnamefont {K.}~\bibnamefont {Kanoda}}, \bibinfo
  {author} {\bibfnamefont {M.}~\bibnamefont {Maesato}}, \ and\ \bibinfo
  {author} {\bibfnamefont {G.}~\bibnamefont {Saito}},\ }\href@noop {}
  {\bibfield  {journal} {\bibinfo  {journal} {Phys. Rev. Lett.},\ }\textbf
  {\bibinfo {volume} {91}},\ \bibinfo {pages} {107001} (\bibinfo {year}
  {2003})}\BibitemShut {NoStop}%
\bibitem [{\citenamefont {Aharen}\ \emph {et~al.}(2010)\citenamefont {Aharen},
  \citenamefont {Greedan}, \citenamefont {Bridges}, \citenamefont {Aczel},
  \citenamefont {Rodriguez}, \citenamefont {MacDougall}, \citenamefont {Luke},
  \citenamefont {Imai}, \citenamefont {Michaelis}, \citenamefont {Kroeker},
  \citenamefont {Zhou}, \citenamefont {Wiebe},\ and\ \citenamefont
  {Cranswick}}]{Aharen}%
  \BibitemOpen
  \bibfield  {author} {\bibinfo {author} {\bibfnamefont {T.}~\bibnamefont
  {Aharen}}, \bibinfo {author} {\bibfnamefont {J.~E.}\ \bibnamefont {Greedan}},
  \bibinfo {author} {\bibfnamefont {C.~A.}\ \bibnamefont {Bridges}}, \bibinfo
  {author} {\bibfnamefont {A.~A.}\ \bibnamefont {Aczel}}, \bibinfo {author}
  {\bibfnamefont {J.}~\bibnamefont {Rodriguez}}, \bibinfo {author}
  {\bibfnamefont {G.}~\bibnamefont {MacDougall}}, \bibinfo {author}
  {\bibfnamefont {G.~M.}\ \bibnamefont {Luke}}, \bibinfo {author}
  {\bibfnamefont {T.}~\bibnamefont {Imai}}, \bibinfo {author} {\bibfnamefont
  {V.~K.}\ \bibnamefont {Michaelis}}, \bibinfo {author} {\bibfnamefont
  {S.}~\bibnamefont {Kroeker}}, \bibinfo {author} {\bibfnamefont
  {H.}~\bibnamefont {Zhou}}, \bibinfo {author} {\bibfnamefont {C.~R.}\
  \bibnamefont {Wiebe}}, \ and\ \bibinfo {author} {\bibfnamefont {L.~M.~D.}\
  \bibnamefont {Cranswick}},\ }\href@noop {} {\bibfield  {journal} {\bibinfo
  {journal} {Phys. Rev. B},\ }\textbf {\bibinfo {volume} {81}},\ \bibinfo
  {pages} {224409} (\bibinfo {year} {2010})}\BibitemShut {NoStop}%
\bibitem [{\citenamefont {de~Vries}\ \emph {et~al.}(2010)\citenamefont
  {de~Vries}, \citenamefont {Mclaughlin},\ and\ \citenamefont
  {Bos}}]{deVries1}%
  \BibitemOpen
  \bibfield  {author} {\bibinfo {author} {\bibfnamefont {M.~A.}\ \bibnamefont
  {de~Vries}}, \bibinfo {author} {\bibfnamefont {A.~C.}\ \bibnamefont
  {Mclaughlin}}, \ and\ \bibinfo {author} {\bibfnamefont {J.-W.~G.}\
  \bibnamefont {Bos}},\ }\href@noop {} {\bibfield  {journal} {\bibinfo
  {journal} {Phys.\ Rev.\ Lett.},\ }\textbf {\bibinfo {volume} {104}},\
  \bibinfo {pages} {177202} (\bibinfo {year} {2010})}\BibitemShut {NoStop}%
\bibitem [{\citenamefont {Mendels}\ \emph {et~al.}(2007)\citenamefont
  {Mendels}, \citenamefont {Bert}, \citenamefont {de~Vries}, \citenamefont
  {Olariu}, \citenamefont {Harrison}, \citenamefont {Duc}, \citenamefont
  {Trombe}, \citenamefont {Lord}, \citenamefont {Amato},\ and\ \citenamefont
  {Baines}}]{Mendels}%
  \BibitemOpen
  \bibfield  {author} {\bibinfo {author} {\bibfnamefont {P.}~\bibnamefont
  {Mendels}}, \bibinfo {author} {\bibfnamefont {F.}~\bibnamefont {Bert}},
  \bibinfo {author} {\bibfnamefont {M.~A.}\ \bibnamefont {de~Vries}}, \bibinfo
  {author} {\bibfnamefont {A.}~\bibnamefont {Olariu}}, \bibinfo {author}
  {\bibfnamefont {A.}~\bibnamefont {Harrison}}, \bibinfo {author}
  {\bibfnamefont {F.}~\bibnamefont {Duc}}, \bibinfo {author} {\bibfnamefont
  {J.~C.}\ \bibnamefont {Trombe}}, \bibinfo {author} {\bibfnamefont {J.~S.}\
  \bibnamefont {Lord}}, \bibinfo {author} {\bibfnamefont {A.}~\bibnamefont
  {Amato}}, \ and\ \bibinfo {author} {\bibfnamefont {C.}~\bibnamefont
  {Baines}},\ }\href@noop {} {\bibfield  {journal} {\bibinfo  {journal} {Phys.\
  Rev.\ Lett.},\ }\textbf {\bibinfo {volume} {98}},\ \bibinfo {pages} {077204}
  (\bibinfo {year} {2007})}\BibitemShut {NoStop}%
\bibitem [{\citenamefont {Lee}\ \emph {et~al.}(2007)\citenamefont {Lee},
  \citenamefont {Kikuchi}, \citenamefont {Qiu}, \citenamefont {Lake},
  \citenamefont {Huang}, \citenamefont {Habicht},\ and\ \citenamefont
  {Kiefer}}]{SHLee}%
  \BibitemOpen
  \bibfield  {author} {\bibinfo {author} {\bibfnamefont {S.~H.}\ \bibnamefont
  {Lee}}, \bibinfo {author} {\bibfnamefont {H.}~\bibnamefont {Kikuchi}},
  \bibinfo {author} {\bibfnamefont {Y.}~\bibnamefont {Qiu}}, \bibinfo {author}
  {\bibfnamefont {B.}~\bibnamefont {Lake}}, \bibinfo {author} {\bibfnamefont
  {Q.}~\bibnamefont {Huang}}, \bibinfo {author} {\bibfnamefont
  {K.}~\bibnamefont {Habicht}}, \ and\ \bibinfo {author} {\bibfnamefont
  {K.}~\bibnamefont {Kiefer}},\ }\href@noop {} {\bibfield  {journal} {\bibinfo
  {journal} {Nature Materials},\ }\textbf {\bibinfo {volume} {6}},\ \bibinfo
  {pages} {853} (\bibinfo {year} {2007})}\BibitemShut {NoStop}%
\bibitem [{\citenamefont {Imai}\ \emph {et~al.}(2008)\citenamefont {Imai},
  \citenamefont {Nytko}, \citenamefont {Bartlett}, \citenamefont {Shores},\
  and\ \citenamefont {Nocera}}]{Imai}%
  \BibitemOpen
  \bibfield  {author} {\bibinfo {author} {\bibfnamefont {T.}~\bibnamefont
  {Imai}}, \bibinfo {author} {\bibfnamefont {E.~A.}\ \bibnamefont {Nytko}},
  \bibinfo {author} {\bibfnamefont {B.~M.}\ \bibnamefont {Bartlett}}, \bibinfo
  {author} {\bibfnamefont {M.~P.}\ \bibnamefont {Shores}}, \ and\ \bibinfo
  {author} {\bibfnamefont {D.~G.}\ \bibnamefont {Nocera}},\ }\href@noop {}
  {\bibfield  {journal} {\bibinfo  {journal} {Phys. Rev. Lett.},\ }\textbf
  {\bibinfo {volume} {100}},\ \bibinfo {pages} {077203} (\bibinfo {year}
  {2008})}\BibitemShut {NoStop}%
\bibitem [{\citenamefont {Olariu}\ \emph {et~al.}(2008)\citenamefont {Olariu},
  \citenamefont {Mendels}, \citenamefont {Bert}, \citenamefont {Duc},
  \citenamefont {Trombe}, \citenamefont {de~Vries},\ and\ \citenamefont
  {Harrison}}]{Olariu}%
  \BibitemOpen
  \bibfield  {author} {\bibinfo {author} {\bibfnamefont {A.}~\bibnamefont
  {Olariu}}, \bibinfo {author} {\bibfnamefont {P.}~\bibnamefont {Mendels}},
  \bibinfo {author} {\bibfnamefont {F.}~\bibnamefont {Bert}}, \bibinfo {author}
  {\bibfnamefont {F.}~\bibnamefont {Duc}}, \bibinfo {author} {\bibfnamefont
  {J.~C.}\ \bibnamefont {Trombe}}, \bibinfo {author} {\bibfnamefont {M.~A.}\
  \bibnamefont {de~Vries}}, \ and\ \bibinfo {author} {\bibfnamefont
  {A.}~\bibnamefont {Harrison}},\ }\href@noop {} {\bibfield  {journal}
  {\bibinfo  {journal} {Phys.\ Rev.\ Lett.},\ }\textbf {\bibinfo {volume}
  {100}},\ \bibinfo {pages} {087202} (\bibinfo {year} {2008})}\BibitemShut
  {NoStop}%
\bibitem [{\citenamefont {de~Vries}\ \emph {et~al.}(2008)\citenamefont
  {de~Vries}, \citenamefont {Kamenev}, \citenamefont {Kockelmann},
  \citenamefont {Sanchez-Benitez},\ and\ \citenamefont {Harrison}}]{deVries2}%
  \BibitemOpen
  \bibfield  {author} {\bibinfo {author} {\bibfnamefont {M.~A.}\ \bibnamefont
  {de~Vries}}, \bibinfo {author} {\bibfnamefont {K.~V.}\ \bibnamefont
  {Kamenev}}, \bibinfo {author} {\bibfnamefont {W.~A.}\ \bibnamefont
  {Kockelmann}}, \bibinfo {author} {\bibfnamefont {J.}~\bibnamefont
  {Sanchez-Benitez}}, \ and\ \bibinfo {author} {\bibfnamefont {A.}~\bibnamefont
  {Harrison}},\ }\href@noop {} {\bibfield  {journal} {\bibinfo  {journal}
  {Phys. Rev. Lett.},\ }\textbf {\bibinfo {volume} {100}},\ \bibinfo {pages}
  {157205} (\bibinfo {year} {2008})}\BibitemShut {NoStop}%
\bibitem [{\citenamefont {Han}\ \emph {et~al.}(2011)\citenamefont {Han},
  \citenamefont {Helton}, \citenamefont {Chu}, \citenamefont {Prodi},
  \citenamefont {Singh}, \citenamefont {Mazzoli}, \citenamefont {Muller},
  \citenamefont {Nocera},\ and\ \citenamefont {Lee}}]{Han}%
  \BibitemOpen
  \bibfield  {author} {\bibinfo {author} {\bibfnamefont {T.~H.}\ \bibnamefont
  {Han}}, \bibinfo {author} {\bibfnamefont {J.~S.}\ \bibnamefont {Helton}},
  \bibinfo {author} {\bibfnamefont {S.}~\bibnamefont {Chu}}, \bibinfo {author}
  {\bibfnamefont {A.}~\bibnamefont {Prodi}}, \bibinfo {author} {\bibfnamefont
  {D.~K.}\ \bibnamefont {Singh}}, \bibinfo {author} {\bibfnamefont
  {C.}~\bibnamefont {Mazzoli}}, \bibinfo {author} {\bibfnamefont
  {P.}~\bibnamefont {Muller}}, \bibinfo {author} {\bibfnamefont {D.~G.}\
  \bibnamefont {Nocera}}, \ and\ \bibinfo {author} {\bibfnamefont {Y.~S.}\
  \bibnamefont {Lee}},\ }\href@noop {} {\bibfield  {journal} {\bibinfo
  {journal} {Phys. Rev. B},\ }\textbf {\bibinfo {volume} {83}},\ \bibinfo
  {pages} {100402R} (\bibinfo {year} {2011})}\BibitemShut {NoStop}%
\bibitem [{\citenamefont {Mendels}\ and\ \citenamefont
  {Bert}(2010)}]{MendelsReview}%
  \BibitemOpen
  \bibfield  {author} {\bibinfo {author} {\bibfnamefont {P.}~\bibnamefont
  {Mendels}}\ and\ \bibinfo {author} {\bibfnamefont {F.}~\bibnamefont {Bert}},\
  }\href@noop {} {\bibfield  {journal} {\bibinfo  {journal} {J. Phys. Soc.
  Jpn.},\ }\textbf {\bibinfo {volume} {79}},\ \bibinfo {pages} {011001}
  (\bibinfo {year} {2010})}\BibitemShut {NoStop}%
\bibitem [{\citenamefont {Elstner}\ and\ \citenamefont
  {Young}(1994)}]{Elstner}%
  \BibitemOpen
  \bibfield  {author} {\bibinfo {author} {\bibfnamefont {N.}~\bibnamefont
  {Elstner}}\ and\ \bibinfo {author} {\bibfnamefont {A.~P.}\ \bibnamefont
  {Young}},\ }\href@noop {} {\bibfield  {journal} {\bibinfo  {journal} {Phys.\
  Rev.\ B},\ }\textbf {\bibinfo {volume} {50}},\ \bibinfo {pages} {6871}
  (\bibinfo {year} {1994})}\BibitemShut {NoStop}%
\bibitem [{\citenamefont {Mila}(1998)}]{Mila}%
  \BibitemOpen
  \bibfield  {author} {\bibinfo {author} {\bibfnamefont {F.}~\bibnamefont
  {Mila}},\ }\href@noop {} {\bibfield  {journal} {\bibinfo  {journal} {Phys.\
  Rev.\ Lett.},\ }\textbf {\bibinfo {volume} {81}},\ \bibinfo {pages} {2356}
  (\bibinfo {year} {1998})}\BibitemShut {NoStop}%
\bibitem [{\citenamefont {Rigol}\ and\ \citenamefont {Singh}(2007)}]{Rigol}%
  \BibitemOpen
  \bibfield  {author} {\bibinfo {author} {\bibfnamefont {M.}~\bibnamefont
  {Rigol}}\ and\ \bibinfo {author} {\bibfnamefont {R.~R.~P.}\ \bibnamefont
  {Singh}},\ }\href@noop {} {\bibfield  {journal} {\bibinfo  {journal} {Phys.\
  Rev.\ Lett.},\ }\textbf {\bibinfo {volume} {98}},\ \bibinfo {pages} {207204}
  (\bibinfo {year} {2007})}\BibitemShut {NoStop}%
\bibitem [{\citenamefont {Misguich}\ and\ \citenamefont
  {Sindzingre}(2007)}]{Misguich}%
  \BibitemOpen
  \bibfield  {author} {\bibinfo {author} {\bibfnamefont {G.}~\bibnamefont
  {Misguich}}\ and\ \bibinfo {author} {\bibfnamefont {P.}~\bibnamefont
  {Sindzingre}},\ }\href@noop {} {\bibfield  {journal} {\bibinfo  {journal}
  {Eur.\ Phys.\ J.\ B},\ }\textbf {\bibinfo {volume} {59}},\ \bibinfo {pages}
  {305} (\bibinfo {year} {2007})}\BibitemShut {NoStop}%
\bibitem [{\citenamefont {Ran}\ \emph {et~al.}(2007)\citenamefont {Ran},
  \citenamefont {Hermele}, \citenamefont {Lee},\ and\ \citenamefont
  {Wen}}]{Ran}%
  \BibitemOpen
  \bibfield  {author} {\bibinfo {author} {\bibfnamefont {Y.}~\bibnamefont
  {Ran}}, \bibinfo {author} {\bibfnamefont {M.}~\bibnamefont {Hermele}},
  \bibinfo {author} {\bibfnamefont {P.~A.}\ \bibnamefont {Lee}}, \ and\
  \bibinfo {author} {\bibfnamefont {X.~G.}\ \bibnamefont {Wen}},\ }\href@noop
  {} {\bibfield  {journal} {\bibinfo  {journal} {Phys.\ Rev.\ Lett.},\ }\textbf
  {\bibinfo {volume} {98}},\ \bibinfo {pages} {117205} (\bibinfo {year}
  {2007})}\BibitemShut {NoStop}%
\bibitem [{\citenamefont {Hermele}\ \emph {et~al.}(2008)\citenamefont
  {Hermele}, \citenamefont {Ran}, \citenamefont {Lee},\ and\ \citenamefont
  {Wen}}]{Hermele}%
  \BibitemOpen
  \bibfield  {author} {\bibinfo {author} {\bibfnamefont {M.}~\bibnamefont
  {Hermele}}, \bibinfo {author} {\bibfnamefont {Y.}~\bibnamefont {Ran}},
  \bibinfo {author} {\bibfnamefont {P.~A.}\ \bibnamefont {Lee}}, \ and\
  \bibinfo {author} {\bibfnamefont {X.~G.}\ \bibnamefont {Wen}},\ }\href@noop
  {} {\bibfield  {journal} {\bibinfo  {journal} {Phys.\ Rev.\ B.},\ }\textbf
  {\bibinfo {volume} {77}},\ \bibinfo {pages} {224413} (\bibinfo {year}
  {2008})}\BibitemShut {NoStop}%
\bibitem [{\citenamefont {Yan}\ \emph {et~al.}(2011)\citenamefont {Yan},
  \citenamefont {Huse},\ and\ \citenamefont {White}}]{Yan}%
  \BibitemOpen
  \bibfield  {author} {\bibinfo {author} {\bibfnamefont {S.}~\bibnamefont
  {Yan}}, \bibinfo {author} {\bibfnamefont {D.~A.}\ \bibnamefont {Huse}}, \
  and\ \bibinfo {author} {\bibfnamefont {S.~R.}\ \bibnamefont {White}},\
  }\href@noop {} {\bibfield  {journal} {\bibinfo  {journal} {Science},\
  }\textbf {\bibinfo {volume} {332}},\ \bibinfo {pages} {1173} (\bibinfo {year}
  {2011})}\BibitemShut {NoStop}%
\bibitem [{\citenamefont {Freedman}\ \emph {et~al.}(2010)\citenamefont
  {Freedman}, \citenamefont {Han}, \citenamefont {Prodi}, \citenamefont
  {Muller}, \citenamefont {Huang}, \citenamefont {Chen}, \citenamefont {Webb},
  \citenamefont {Lee}, \citenamefont {McQueen},\ and\ \citenamefont
  {Nocera}}]{Freedman}%
  \BibitemOpen
  \bibfield  {author} {\bibinfo {author} {\bibfnamefont {D.~E.}\ \bibnamefont
  {Freedman}}, \bibinfo {author} {\bibfnamefont {T.~H.}\ \bibnamefont {Han}},
  \bibinfo {author} {\bibfnamefont {A.}~\bibnamefont {Prodi}}, \bibinfo
  {author} {\bibfnamefont {P.}~\bibnamefont {Muller}}, \bibinfo {author}
  {\bibfnamefont {Q.~Z.}\ \bibnamefont {Huang}}, \bibinfo {author}
  {\bibfnamefont {Y.~S.}\ \bibnamefont {Chen}}, \bibinfo {author}
  {\bibfnamefont {S.~M.}\ \bibnamefont {Webb}}, \bibinfo {author}
  {\bibfnamefont {Y.~S.}\ \bibnamefont {Lee}}, \bibinfo {author} {\bibfnamefont
  {T.~M.}\ \bibnamefont {McQueen}}, \ and\ \bibinfo {author} {\bibfnamefont
  {D.~G.}\ \bibnamefont {Nocera}},\ }\href@noop {} {\bibfield  {journal}
  {\bibinfo  {journal} {J.\ Am.\ Chem.\ Soc.},\ }\textbf {\bibinfo {volume}
  {132}},\ \bibinfo {pages} {16185} (\bibinfo {year} {2010})}\BibitemShut
  {NoStop}%
\bibitem [{\citenamefont {Azuma}\ \emph {et~al.}(1994)\citenamefont {Azuma},
  \citenamefont {Hiroi}, \citenamefont {Takano}, \citenamefont {Ishida},\ and\
  \citenamefont {Kitaoka}}]{Azuma}%
  \BibitemOpen
  \bibfield  {author} {\bibinfo {author} {\bibfnamefont {M.}~\bibnamefont
  {Azuma}}, \bibinfo {author} {\bibfnamefont {Z.}~\bibnamefont {Hiroi}},
  \bibinfo {author} {\bibfnamefont {M.}~\bibnamefont {Takano}}, \bibinfo
  {author} {\bibfnamefont {K.}~\bibnamefont {Ishida}}, \ and\ \bibinfo {author}
  {\bibfnamefont {Y.}~\bibnamefont {Kitaoka}},\ }\href@noop {} {\bibfield
  {journal} {\bibinfo  {journal} {Phys. Rev. Lett.},\ }\textbf {\bibinfo
  {volume} {73}},\ \bibinfo {pages} {3463} (\bibinfo {year}
  {1994})}\BibitemShut {NoStop}%
\bibitem [{\citenamefont {Lee}\ and\ \citenamefont {Lee}(2005)}]{SSLee}%
  \BibitemOpen
  \bibfield  {author} {\bibinfo {author} {\bibfnamefont {S.~S.}\ \bibnamefont
  {Lee}}\ and\ \bibinfo {author} {\bibfnamefont {P.~A.}\ \bibnamefont {Lee}},\
  }\href@noop {} {\bibfield  {journal} {\bibinfo  {journal} {Phys.\ Rev.\
  Lett.},\ }\textbf {\bibinfo {volume} {95}},\ \bibinfo {pages} {036403}
  (\bibinfo {year} {2005})}\BibitemShut {NoStop}%
\end{thebibliography}



%

\end{document}